\documentclass[journal]{IEEEtran}
\usepackage{}

\usepackage{amsfonts}
\usepackage{amssymb}
\usepackage{amsmath}
\usepackage{mathrsfs}
\usepackage{amsmath,amssymb}
\usepackage{amsmath}
\usepackage{graphicx}
\usepackage{amsmath,amssymb,amsthm}
\usepackage{leftidx}
\usepackage{extarrows}
\usepackage{caption}
\usepackage{enumerate}
\usepackage{graphics}
\usepackage{subfigure}
\usepackage{float}
\usepackage{pict2e}
\usepackage{tikz}
\usepackage{cases}
\usepackage[numbers,sort&compress]{natbib}

\ifCLASSINFOpdf
\else
\fi
\begin{document}
%
\title{A perspective on Attitude Control Issues and Techniques}
%
%
%

\author{Dandan Zhang, Xin Jin, Hongye Su
\thanks{This work was partially supported by Science Fund for Creative Research Group of the
National Natural Science Foundation of China (Grant NO.61621002),
by Zhejiang Key R\&D Program (Grant NO. 2021C01198, 2022C01035) and by China Postdoctoral Science Foundation under Grant 2021TQ0107.
}
\thanks{D. Zhang and H. Su are with the
State Key Laboratory of Industrial Control Technology, Institute of
Cyber-Systems and Control, Zhejiang University, Hangzhou 310027,
China (e-mail: zdandan3@126.com; hysu@iipc.zju.edu.cn, corresponding
author).

X. Jin is with the Key Laboratory of Smart Manufacturing in Energy Chemical Process,
Ministry of Education, East China University of Science and Technology, Shanghai 200237, China
(e-mail: $\mathrm{jx}_{-}$9810@163.com).}
}

\maketitle

\begin{abstract}
This paper reviews the attitude control problems for rigid-body systems, starting from the attitude representation for rigid body kinematics. Highly redundant rotation matrix defines the attitude orientation globally and uniquely by 9 parameters, which is the most fundamental one, without any singularities; minimum 3-parameter Euler angles or (modified) Rodrigues parameters define the attitude orientation neither globally nor uniquely, but the former exhibits kinematical singularity and Gimbal lock, while the latter two exhibit geometrical singularity; once-redundant axis-angle or unit quaternion globally define the attitude rotation but not uniquely using 4 parameters, but the former is not appropriate to define very small or very large rotations, while the latter shows unwinding phenomenon despite of the reduced computation burden. In addition, we explore the relationships among those attitude representations, including the connections among Gimbal lock, unwinding phenomenon and a nowhere dense set of zero Lebesgue measure. Based on attitude representations, we analyze different attitude control laws, almost global control and global attitude control, nominal and general robustness, as well as the technique tools.

\end{abstract}


%
\IEEEpeerreviewmaketitle


\section{Overview}


\IEEEPARstart{R}{igid}-body attitude control problems have received considerable research interest since the 1950s \cite{10.5[5]} and have been a long-standing staple in the nonlinear control along with a vast literature spanning many decades \cite{8.1[1]}, \cite{10.7[1]}, \cite{8.1[3]}, \cite{10.7}, \cite{18abd}, \cite{2010mengziyangauto}, \cite{du2016tac}, \cite{zddtcsI}, \cite{zddtcns}, \cite{jinxinauto}, \cite{jinxintcst}, with various applications in aerospace, underwater vehicles, marine engineering and robotics \cite{43abd}, \cite{renwei2008shu}, \cite{2018liutao-auto}, \cite{Dobook}.
For the investigation convenience, it is usually assumed that the states of the rigid body can be placed into an abstract manifold with appropriate dimensions, with the one-to-one corresponding relations of points between them \cite{2012tac[4]}.
Under this correspondence, the attitude dynamics of the rigid body give rise to a dynamical system on the boundaryless compact manifold $SO(3)$, i.e., the state space of attitude rotation matrices \cite{2011tac}, \cite{pedro2017auto}.
In fact, it is precisely because the topological property without contractibility of $SO(3)$, resulting from the compact property without boundary (see in \cite[Ex. 2.4.6]{8.1[9]}), that $SO(3)$ is not diffeomorphic to any Euclidean space. That is, for a differential equation on the attitude dynamics/kinematics with a locally Lipschitz right-hand side, the basin of attraction of its asymptotically stable equilibrium point is not contractible.

On the attitude configuration space $SO(3)$, the rigid-body attitude is usually parameterized for exploiting the redundancies
in the rotation-matrix descriptions. According to the attitude parametrization, attitude control strategies can be
categorized by choosing the Euler-angles parametrization (e.g., roll, yaw and pitch), exponential coordinates, the (modified) Rodrigues parametrization \cite{18abd} and the unit quaternion parametrization \cite{13-14}, etc.
However, different parameterizations lead to further topological difficulties. For example, considering that no globally nonsingular three-parameter representation exists, the above former three parametrizations would consequently cause the singularity issue \cite{12[14]}, \cite{10.7}.
This can seriously constrain admissible
attitudes and further complicates the path planning.
This obstacle on the allowable attitude motions has nothing to do with the
physically admissible attitude motions, while it is for purely
mathematical reasons, i.e., it is an inherent obstacle.
As a globally nonsingular parametrization (e.g., the covering map from
$\mathbb{S}^{3}$ to $SO(3)$ is everywhere a local diffeomorphism), the unit quaternion parametrization is often employed evolving on the three-dimensional unit sphere $\mathbb{S}^{3}$.
However, the well-known Rodrigues mapping from $\mathbb{S}^{3}$ to $SO(3)$ implies that two antipodal unit quaternions correspond to the same attitude in $SO(3)$. Corresponding to the above described topological constraints on $SO(3)$, the attitude control law designed on $\mathbb{S}^{3}$ is to stabilize two disconnected sets of quaternions consistent with the same physical attitude on $SO(3)$ \cite{2011tac}, \cite{2012tac}, \cite{2013tac}, \cite{13-14}. Without considering the double covering problem \cite{10.7}, the designed controller may induce the unwinding phenomenon where the rigid body will unnecessarily make a full rotation \cite{2012tac[4]}. To avoid the unwinding phenomenon, a dynamic mechanism is necessary to
resolve the ambiguity, i.e., which quaternion should be chosen for feedback, when an inconsistent quaternion-based feedback can be used simultaneously. More specifically, a hysteretic-based switching scheme is considered in terms of the aforementioned binary logic variable, to select which pole of $\mathbb{S}^{3}$ to regulate in a hysteretic fashion, see \cite{2011tac} for more detailed analysis. Later, this unwinding-free feedback strategy has been further applied into \cite{2012tac}, \cite{2013tac} and \cite{13-14}, which manages a tradeoff between the robustness to arbitrarily small measurement disturbances and a small amount of hysteresis-induced inefficiency caused by the unwinding phenomenon. Note that, in \cite{18abd}, \cite{20abd} and \cite{43abd} for the synchronization problem, the unwinding phenomenon has been solved by restricting the control gain, and the synchronization can be realized if and only if the imaginary part of a quaternion is zero. Coincidentally, in the later work \cite{2014duauto} on the synchronization issue, the authors have made that point emphatically in the reply to the comments on the quaternion ambiguity involved in \cite{2014duauto} made by Rezaee and Abdollahi \cite{com2014}, by assuming that the scale part of quaternion for each spacecraft is
always positive. Therefore, the quaternion ambiguity should be considered in the hybrid feedback strategy for the attitude investigation of rigid bodies, to exclude the unwinding phenomenon.

Compared with other global (non-singular) attitude representations (i.e., rotation matrix also called direction cosine matrix, abbr. DCM) and the axis-angle parameterization, the computational burden can be greatly reduced by making use of unit quaternions \cite[pg. 322]{shs1.1Ehsan2019[9]}. As the minimal globally nonsingular parametrization, unit quaternions are often used to parametrize rotation matrices on $SO(3)$; meanwhile, however, the unwinding phenomenon is introduced due to the double covering transformation from $SO(3)$ to state space $\mathbb{S}^{3}$ of unit quaternions \cite{2012tac[4]}, causing the rigid body to unnecessarily make a full rotation from the unstable saddle point to another stable equilibrium point
\cite{10.7}, \cite{chat2009tac}. Nevertheless, the topology constraint on $SO(3)$ never stops the researchers from using unit quaternions to design feedback control algorithms for attitude control \cite{du2016tac}, \cite{13-14}. To solve the topology constraint and realize the robust global attitude control, the design of the quaternion-based torque control law applies a binary logic variable \cite{2011tac, 2012tac}, where the hysteresis half-width arranges a tradeoff between the hysteresis-induced inefficiency for avoiding unwinding phenomenon and the robustness to measurement noise.
Hence, the salient feature on the attitude control problem leads to a captivating difficulty stemming from the well-known topological constraints of the rigid-body state space $SO(3)$ on rotation matrices: It is impossible to find any continuous feedback control law so that all equilibrium points of $SO(3)$ can be globally asymptotically stable, since there must be one other unstable equilibrium point in any smooth
vector field on $SO(3)$ if an attracting equilibrium point exists, as analyzed in \cite{2012tac}, \cite{2013tac}, \cite{36abd} and \cite{13-14}.

With continuous state-feedback control laws designed on the Lie group $SO(3)$ or the Lie group diffeomorphic to $SO(3)\times \mathbb{R}^{3}$ (like \cite{chat2009tac}), the best results that one can achieve are at most almost global stability (i.e., the attraction domain of the equilibrium is open and dense), where the desired attitude can be stabilized from any initial attitude, necessarily excluding a nowhere dense set of zero Lebesgue measure which is attracted to the antipode of the stabilized desired orientation. But yet, for the same reason, there exist no periodic or discontinuous feedback that can simultaneously stabilize a particular attitude robustly and globally: Any periodic or discontinuous feedback will create a rigid decision boundary, and it can be hijacked by any small measurement disturbance so that the real-time attitude attitude is far from the reference one, as verified in \cite{2011tac}. To compensate for the topological constraints on $SO(3)$ by simultaneously achieving the global and robust attitude control performance, many authors employ the hybrid feedback control strategy \cite{2011tac}, \cite{2012tac}, \cite{2013tac}, \cite{13-14}, \cite{8.1}, \cite{36abd}, \cite{15abd}, \cite{16abd}. More specifically, in \cite{2011tac}, the authors propose several hybrid control strategies based on the binary logic variable $h\in \{-1,1\}$ in terms of the \emph{sign} function, and verify the most appropriate one by considering the well-posed property of hybrid dynamic theory.
Later, the authors in \cite{2012tac}, \cite{2013tac} and \cite{13-14} employ the logic-variable based hybrid strategy to different investigations including the attitude stabilization/consensus/synchronizaiton, etc.
Differently, in \cite{8.1}, \cite{36abd}, \cite{15abd} and \cite{16abd}, the authors consider the inherent passivity properties of attitude dynamics to design an appropriate error function on $SO(3)$, and further establish an artificial potential functions. It is verified that, by coordinating a family of synergistic potential functions in a hybrid strategy with a hysteresis-based
switching mechanism by selecting an appropriate control law corresponding to the minimal potential function, the topological obstructions can be overcome. Based on the \cite{8.1}, the authors in \cite{36abd}, \cite{15abd} and \cite{16abd} propose the exp-synergistic potential functions on $SO(3)$, which are applied to velocity-free hybrid attitude stabilization.

\begin{table*}[tbp]
\centering  
\begin{tabular}{|c|c|c|c|}
\hline
Attitude Representation & Globally represent  & Uniquely represent & Number of parameters \\ \hline

Euler angles (kinematically singular)&  No  & No & 3 \\ \hline


(Modified) Rodrigues parameters (geometrically singular)    & No & No& 3  \\ \hline

Quaternions& Yes & No& 4\\ \hline
Axis-angle & Yes & No & 4\\  \hline
Rotation matrix & Yes & Yes & 9\\  \hline
\end{tabular}
 \caption{Properties of attitude representations I.}\label{unique}
\end{table*}

This paper serves as a roadmap for the theoretical development of attitude control, including the following aspects:\\
\indent 1) A class of attitude representations, i.e., 9-parameter rotation matrix (globally and uniquely, no singularities), 3-parameter Euler angles (neither globally nor uniquely, kinematical singularity and Gimbal lock) and (modified) Rodrigues parameters (neither globally nor uniquely, geometrical singularity), 4-parameter axis-angle (globally but not uniquely) and unit quaternion (globally but not uniquely, unwinding phenomenon, reduced computation burden).\\
\indent 2) The relationships among those attitude representations, especially the connections among Gimbal lock, unwinding phenomenon and a nowhere dense set of zero Lebesgue measure, i.e., Gimbal lock (i.e., when the plane's head is straight up or straight down) includes a nowhere dense set of zero Lebesgue measure, not including the two antipodal equilibrium points. The unwinding phenomenon is caused when we ignore the two antipodal equilibrium points, so even when we remove the Gimbal lock problem using unit quaternions, then unwinding phenomenon could still occur.\\
\indent 3) Attitude control laws for the (almost) global control, including unit quaternion based control laws (continuous, discontinuous and hybrid cases) and DCM based control laws. Generally, the hybrid case focuses on the robust global attitude control by overcoming the topology constraints of boundaryless compact manifold $SO(3)$, such as a family of configuration error functions or synergistic potential functions, or binary logic variable based hybrid control. The continuous and discontinuous cases try to realize the global attitude control by adopting certain methods, for example modifying control laws over a set of zero Lebesgue measure or restricting the rotation angle for the global control. \\
\indent 4) The nominal robustness with respect to small perturbations (including measurement noises/disturbances) for avoiding chattering, and general robustness against bounded \emph{unstructured} uncertainties in both the translational dynamics and the rotational dynamics in terms of the presented control laws, where continuous and hybrid control laws are generally nominally robust, not including the purely discontinuous control law.\\
\indent 5) The technique tools used to prove the (almost) global attitude control and the robustness. The LaSalle's/hybrid invariance principle is better applied to aysmptotic stability for time-invariant or periodic equation, not including the time-varying (i.e, non-autonomous) systems. Other than LaSalle's/hybrid invariance principle, cross-term-added Lyapunov function or Barbalat's lemma are feasible for time-varying case. Comparison principle can be applied into non-autonomous/autonomous case, while Matrosov' theorem can be applied in all circumstance, i.e., the time-(in)variant systems, nonperiodic/time-dependent systems, etc. Non-autonomous systems could be dealt with Barbalat's lemma, Matrosov's theorem, comparison principle, etc.

The paper is organized as follows: Section \ref{section3.0} presents attitude representations for rigid body kinematics, including the connections among Gimbal lock, unwinding phenomenon and a nowhere dense set of zero Lebesgue measure. Section \ref{section4.0} presents attitude control laws, including unit quaternion based control laws (continuous, discontinuous and hybrid cases), DCM based control laws, analysis on (almost) global control, analysis on robustness and technique tools. Section \ref{section6} concludes the paper.

\section{Attitude Representation for Rigid Body Kinematics}\label{section3.0}

The first step for investigating attitude control problems of rigid-body systems is to choose a best description for rigid body orientation, which is a very fundamental and important topic. Indeed, choosing a good attitude coordinates could greatly facilitate investigations, i.e., simplifying mathematical calculations and analysis, avoiding geometrical/mathematical singularities or strictly nonlinear kinematic differential equations (KDEs). Generally, in order to choose a good attitude description, one need to consider the following four truths about rigid body attitude coordinates:\\
\indent 1) The minimum parameter used to describe the relative attitude orientation (i.e., angular displacement)
between two reference frames is three.\\
\indent 2) For any 3-parameter set of attitude coordinates, there exists at least one geometrical orientation so that the coordinates are singular, i.e., at least two coordinates are not unique.\\
\indent 3) Corresponding to any geometric singularity, the kinematic
differential equations (KDEs) are also singular.\\
\indent 4) A solution to avoid geometric singularities is to regularize the three-parameter representation: Once-redundant redundant four or more coordinates contain no any geometric/kinematicla singularity, which are universally determined.\\
According to the above truths, we analyze the following attitude representation for rigid-body kinematics.

\subsection{Rotation matrix}\label{section3.1}

In the three-dimensional space, the attitude of a rigid-body system denotes the relative rotation from the body frame $\mathcal {F}_{b}$ to the inertial reference frame $\mathcal {F}_{o}$, called rotation matrix. Rotation matrix is also called direction cosine matrix (DCM), since the axes of $\mathcal {F}_{b}$ and $\mathcal {F}_{o}$ are usually unit vectors, where only cosine terms are involved in the rotation matrix, see \cite[pg. 80-86]{2019Ehsan[27]} for more details. According to \cite[pg. 17]{abd2013shu}, the attitude of a rigid-body system can be described by a $3\times3$ orthogonal rotation matrix unitary determinant, i.e., an element of a Lie group:
\begin{equation*}\label{rm1}
\begin{split}
SO(3)=\{R\in \mathbb{R}^{3\times 3}: R^{T}R=I_{3\times 3}, \hbox{det} (R)=1\},\\
\end{split}
\end{equation*}
where $I$ denotes the $3\times3$ identity element and $\hbox{det} (R)$ is the determinant of matrix $R$.

Take a simple example to show how to use the rotation matrix \cite[pg. 17]{abd2013shu}. Let $R\in SO(3)$ be the rotation from $\mathcal {F}_{b}$ to $\mathcal {F}_{o}$, and given $X_{b}$ as the coordinates of a vector in $\mathcal {F}_{b}$, then the coordinates $X_{o}$ of this same vector in $\mathcal {F}_{o}$ can be denoted as
\begin{equation*}\label{rm2}
\begin{split}
X_{o}=RX_{b}.
\end{split}
\end{equation*}
In fact, the above property also holds for several frames using the composition of rotations, i.e., the noncommutative multiplication
of rotation matrices in terms of those frames.

The definition of rotation matrix shows that the rotation matrix is the most fundamental attitude representation, which can globally and uniquely denote any orientation as shown in Table \ref{unique}. However, the 9-parameter representation also exhibits high redundancy for describing a relative orientation. As shown in Table \ref{unique}, compared with the 3-parameter representation, there are six extra parameters causing redundancy due to the orthogonality condition $R^{T}R=I_{3\times 3}$. In practice, we usually apply the less redundant representation, rather than the 9-parameter DCM.

But at the cost of high redundancy, what we have to admit is that the DCM has a biggest asset, i.e., easily transforming vectors between any two reference frames. Furthermore, given any instant, the instantaneous DCM could be obtained via a rigorously linear differential equation as shown in \cite[pg. 85-86]{2019Ehsan[27]}. So linearity and universal applicability are major advantages of the kinematic differential equations for DCM, and there is no any geometric/kinematical singularities occurring in the kinematic differential equations using DCM.

\begin{figure}[!htb]
\centering
\includegraphics[width=3in,height=2in]{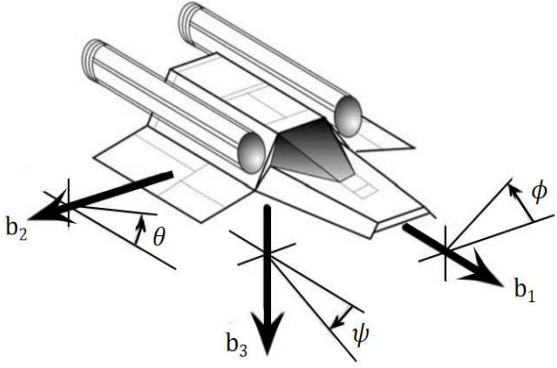}
\caption{\cite[pg. 87]{2019Ehsan[27]} Yaw $\psi$, pitch $\textcolor[rgb]{0.00,0.00,0.00}{\theta}$, and roll $\phi$ Euler angles with respect to body-fixed axes $\{\vec{b}\}=\{b_{1}, b_{2},b_{3}\}$.}\label{rpy}
\end{figure}

\subsection{Euler angles: kinematically singular}\label{section3.2}

The Euler angles are defined to describe the orientation of a rigid body, including three angles with respect to a fixed
coordinate system. Both kinds of sequences symmetric Proper Euler angles (i.e., $z-y-z$, $x-z-x$, $y-x-y$, $z-x-z$, $x-y-x$, $y-z-y$, also called longitude of the ascending node denoted as $\tilde{\psi}$, inclination denoted as $\textcolor[rgb]{0.00,0.00,0.00}{\tilde{\theta}}$, and argument of the perihelion denoted as $\tilde{\phi}$) and asymmetric Tait-Bryan angles ($x-z-y$, $z-y-x$, $y-x-z$, $x-y-z$, $y-z-x$, $z-x-y$, also called yaw-pitch-roll angles) are called ``Euler angles''.

Take (yaw, pitch, roll):=($\psi$, $\textcolor[rgb]{0.00,0.00,0.00}{\theta}$, $\phi$) for example about the sequentially
displaced body-fixed axes $\{\vec{b}\}=\{b_{1}, b_{2},b_{3}\}$, as shown in Fig. \ref{rpy}. Successive multiplication (usually three elemental rotation matrices) of rotation matrices can derive the transformation mapping the vector ($\psi$, $\textcolor[rgb]{0.00,0.00,0.00}{\theta}$, $\phi$) into its corresponding rotation matrix as shown in \cite[pg.18]{abd2013shu}, which shows the connections between rotation matrix and Euler angles, i.e., $(\psi, \textcolor[rgb]{0.00,0.00,0.00}{\theta}, \phi)\longmapsto b_{3}(\psi)b_{2}(\textcolor[rgb]{0.00,0.00,0.00}{\theta})b_{1}(\phi)$, implying $R_{\vec{b}}(\psi, \textcolor[rgb]{0.00,0.00,0.00}{\theta}, \phi)=b_{3}(\psi)b_{2}(\textcolor[rgb]{0.00,0.00,0.00}{\theta})b_{1}(\phi)$, i.e., rotated about axis $b_{3}$ by the yaw
angle $\psi$, axis $b_{2}$ by the pitch angle $\textcolor[rgb]{0.00,0.00,0.00}{\theta}$ and axis $b_{1}$ by the roll angle $\phi$, respectively.

\begin{figure*}
\centering\subfigure[]{
\begin{minipage}[t]{0.33\textwidth}
\label{fig:subfig:c} 
\includegraphics[width=2.2in,height=1.7in]{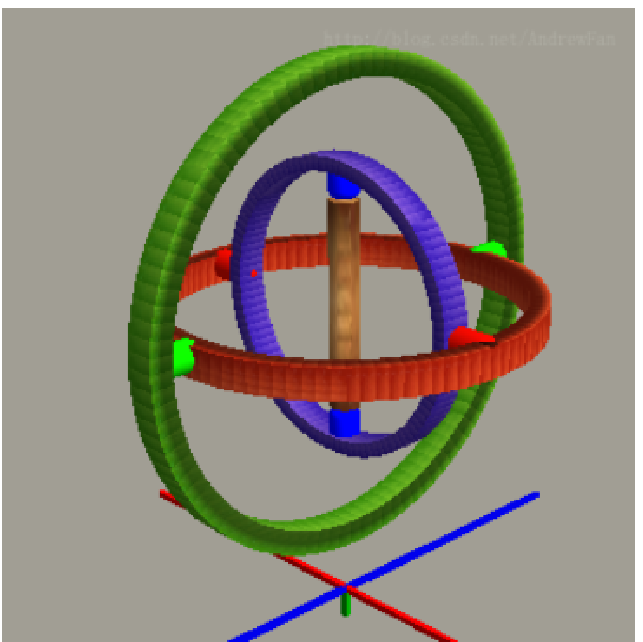}
\end{minipage}}
\centering\subfigure[]{
\begin{minipage}[t]{0.33\textwidth}
\label{fig:subfig:c} 
\includegraphics[width=2.2in,height=1.7in]{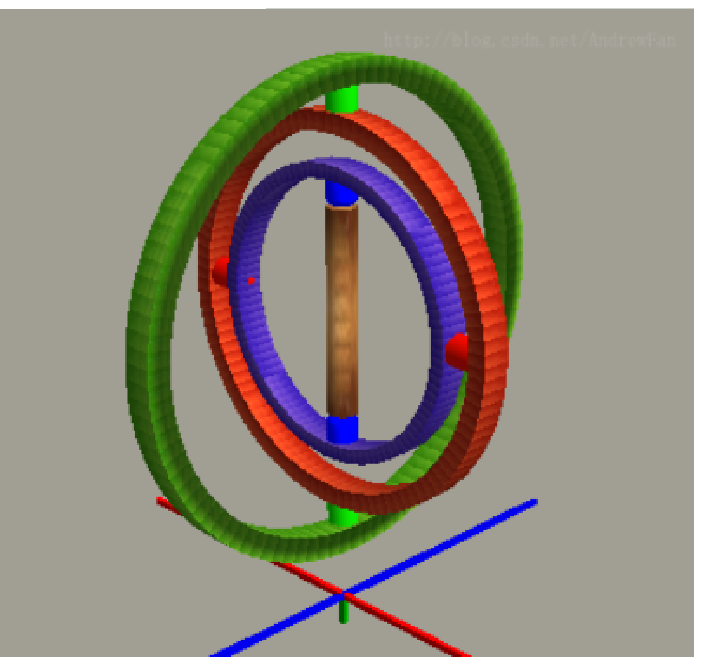}
\end{minipage}}
\caption{The rotor of the gyroscope$^{2}$. (a) The rotor of the gyroscope is kept in balance: Red gimbal ring rotates along $x$-axis (red line) to realize the pitch motion; Blue gimbal ring rotates along $y$-axis (green line) to realize the yaw motion; Green gimbal ring rotates along $z$-axis (blue line) to realize the roll motion. (b) The the rotor of the gyroscope is out of balance: Gimbal lock, i.e., two out of the three gimbals are in the same plane, so that one degree of freedom is lost.}\label{GB}
\end{figure*}

Any orientation can be achieved by composing three elemental rotations, starting from a known standard orientation. Equivalently, any
rotation matrix $R$ can be decomposed as a product of three elemental rotation matrices. Euler angle is a minimal 3-parameter representation of attitude as shown in Table \ref{unique}. The 3-parameter sets of (modified) Rodrigues parameters can be regarded as certain embedded subsets of Euclidean space $\mathbb{R}^{3}$, so that one can apply the analysis methods suited to $\mathbb{R}^{3}$. However, it is not globally defined for the transformation from their time rates of change to the angular velocity vector. In other words, the time derivatives of the Euler angles could not represent every possible angular velocity, causing mathematical singularity of the Euler angle kinematic differential equation (kinematical singularity, i.e., two angles are not uniquely defined). In fact, except rotation matrix, all parameterizations fail to uniquely represent the set of attitudes, and Euler angles are no exception. The angles are uniquely determined except for the singular case that two out of the three gimbals are identical, i.e. two axes have the same or opposite directions, usually called Gimbal lock, as shown in Fig. \ref{GB}. Also as analyzed in \cite[pg. 86-95]{2019Ehsan[27]}, the geometric singularity occurs when the \emph{pitch} is rotated $90^{\circ}$ up or down ($\pm 90^{\circ}$) in pitch-yaw-roll angles, or when an \emph{inclination} angle of $0^{\circ}$ or $180^{\circ}$ in the symmetric Proper Euler angles\footnotemark[1]. So Euler angle representation encounters a geometric singularity at certain specific values of the second Euler rotation angle (pitch angle and/or inclination angle) only, i.e., when the first and third angles are measured in the same plane, however, never encountering a singularity in the first or third angles. Actually, Gimbal lock would cause disastrous results if the rigid-body is in a steep ascent or dive, in a gimbal-based aerospace inertial navigation system.

\footnotetext [1]{The geometric singularity is when the plane's head is straight up or straight down as shown in Fig. \ref{GB}(b), corresponding to ``$\hbox{pitch}=\pm 90^{\circ}$'' or ``$\hbox{inclination}=0^{\circ}~or~180^{\circ}$'' using different sets of Euler angles, called Gimbal lock. More specifically, a singularity (Gimbal lock) occur only at certain specific values of the second rotation angle, never resulting from the first and third rotation angles (the two angles never lead to any singularity). That is, the singularity of attitude (Gimbal lock) may include many combinations of attitude angles, even if the second angle is fixed, which will be analyzed later.}

Therefore, in both kinds of sequences symmetric Proper Euler angles and asymmetric Tait-Bryan angles, a rigid-body is not allowed to rotate more than $90^{\circ}$ away from the singularity, so that they are limited to describe large (in particular arbitrary) rotations. Moreover, despite the kinematic differential equations of Euler angle could be linearized, however, the linearized one could only be valid for relatively small rotations. If not linearized, their fairly nonlinear kinematic differential equations will contain many computationally intensive trigonometric functions. Hence, Euler angles with the minimum number 3 are not always convenient as analyzed in a series of work
\cite[pg. 86-95]{2019Ehsan[27]} or \cite{meyer1966, 10.7[1]}, especially the symmetric Proper Euler angles, since the small departure rotations with vary small angles are always hovering around the singular attitude $0^{\circ}$ (very close to the singularity). Still and all, the 3-parameter Euler angle representation is a compact attitude description with a easily visualized coordinates.


\subsection{Axis-angle}\label{section3.3}


Axis-angle representation is about a given normalized vector by a given rotation angle, which are mapped on a rotation matrix, so the axis-angle representation is a 4-parameter representation of attitude as shown in Table \ref{unique}.

Define $\vec{k}\in \mathbb{R}^{3}$ as an arbitrary unit-length vector and $\textcolor[rgb]{0.00,0.00,0.00}{\Phi}$ as the rotation angle about $\vec{k}$. So we derive the corresponding rotation matrix $\mathbf{R}(\textcolor[rgb]{0.00,0.00,0.00}{\Phi},\vec{k})\in SO(3)$ \cite[pg. 18]{abd2013shu},
\begin{equation}\label{Aa1}
\begin{split}
\mathbf{R}(\Phi, \vec{k})=I_{3}-\sin(\textcolor[rgb]{0.00,0.00,0.00}{\Phi})S(\vec{k})+(1-\cos(\textcolor[rgb]{0.00,0.00,0.00}{\Phi}))S^{2}(\vec{k}),
\end{split}
\end{equation}
where $I_{3}$ denotes the $3\times3$ identity element, and $S(\vec{k})$ is defined as for $\vec{k}=(k_{1}, k_{2}, k_{3})$
\begin{equation*}\label{Aa2}
\begin{split}
S(\vec{k})=\left[\begin{array}{ccc}
       0 & -k_{3}& k_{2}\\
       k_{3}& 0 & -k_{1}\\
       -k_{2}& k_{1}& 0
     \end{array}\right].
\end{split}
\end{equation*}
That is, $S(\cdot)$ denotes the skew symmetric matrix, so the cross product of any two vectors $X,Y \in \mathbb{R}^{3}$ can be expressed by the matrix multiplication, i.e., $X\times Y=S(X)Y$.

Therefore, it can be checked from (\ref{Aa1}) that the attitude matrix can be defined by axis-angle representation, i.e., a given set of parameters $\{\textcolor[rgb]{0.00,0.00,0.00}{\Phi}, \vec{k}\}$. For (\ref{Aa1}), it can be easily checked that $\mathbf{R}(\textcolor[rgb]{0.00,0.00,0.00}{\Phi},\vec{k})=\mathbf{R}(-\textcolor[rgb]{0.00,0.00,0.00}{\Phi},-\vec{k})$ since $S(-\vec{k})=-S(\vec{k})$. More specifically, one can get a unique rotation matrix from a given axis-angle parameters $\{\textcolor[rgb]{0.00,0.00,0.00}{\Phi}, \vec{k}\}$, but not vise versa, since there exist at least two antipodal sets of axis-angle parameters $(\textcolor[rgb]{0.00,0.00,0.00}{\Phi},\vec{k})$ and $(-\textcolor[rgb]{0.00,0.00,0.00}{\Phi},-\vec{k})$ for a given rotation $\mathbf{R}(\textcolor[rgb]{0.00,0.00,0.00}{\Phi},\vec{k})$. Indeed, for the case $\mathbf{R}(\textcolor[rgb]{0.00,0.00,0.00}{\Phi},\vec{k})=I_{3}$, one can find a series of axis-angle parameters $\{2n \pi, \vec{k}\}$ for any $n\in \mathbb{Z}$ and any unit-length vector $\vec{k}\in \mathbb{R}^{3}$.

As analyzed in \cite[pg. 102]{2019Ehsan[27]}, the kinematic differential equation using axis-angle contains a $\frac{0}{0}$-type mathematical singularity when $\textcolor[rgb]{0.00,0.00,0.00}{\Phi}=0$ (i.e., zero rotation), so the axis-angle is not well suited in those cases with small motion feedback control. In addition, the axis-angle is not well suited for the large rotation either, since the axis-angle based mathematical expression contains not only polynomial fractions of degrees up to three but also trigonometric functions, which makes computations more complicated. As a result, axis-angle representation is less attractive to describe small/large arbitrary rotations, compared with other attitude representations.

\subsection{Quaternions (i.e., Euler symmetric parameters)}\label{section3.4}

As shown in Table \ref{unique}, unit quaternions and the axis-angle representation are both 4-parameter representation of attitude. In fact, unit quaternions are often regarded as an axis-angle representation, i.e., $q=(\cos \frac{\Phi}{2},\vec{k}\sin \frac{\Phi}{2})\in \mathbb{S}^{3}$ for given axis-angle parameter $\{\textcolor[rgb]{0.00,0.00,0.00}{\Phi}, \vec{k}\}$, where $\mathbb{S}^{n}:=\{x\in \mathbb{R}\times\mathbb{R}^{n}: x^{T}x =1\}$ is the $n$-dimensional unit sphere embedded in $\mathbb{R}^{n+1}$


According to the Rodrigues formula $\mathcal {R}(-q)=\mathcal {R}(q)=I+2\eta S(\epsilon)+2S^{2}(\epsilon)$
in terms of $\mathcal {R}: \mathbb{S}^{3}\rightarrow SO(3)$, where unit quaternion
$q=(\eta, \epsilon)\in\mathbb{S}^{3}$, every element of $SO(3)$ can be parameterized by two antipodal unit quaternions in $\mathbb{S}^{3}$.
For any two unit quaternions $q_{1}$ and $q_{2}$ with $q_{i}=(\eta_{i}, \epsilon_{i})\in \mathbb{S}^{3}$ for $i=\{1,2\}$, one has
\begin{equation*}\label{t1t}
\begin{split}
q_{1}\otimes q_{2}=\left[ \begin{array}{c}
                              \eta_{1}\eta_{2}- \epsilon^{T}_{1}\epsilon_{2} \\
                              \eta_{1}\epsilon_{2}+ \eta_{2}\epsilon_{1}+S(\epsilon_{1})\epsilon_{2}
                            \end{array}
\right].
\end{split}
\end{equation*}
This implies that each unit quaternion $q_{i}\in\mathbb{S}^{3}$ has an inverse $q^{-1}_{i}=(\eta_{i}, -\epsilon_{i})$.

\begin{figure}[!htb]
\centering
\includegraphics[width=2.8in,height=2.2in]{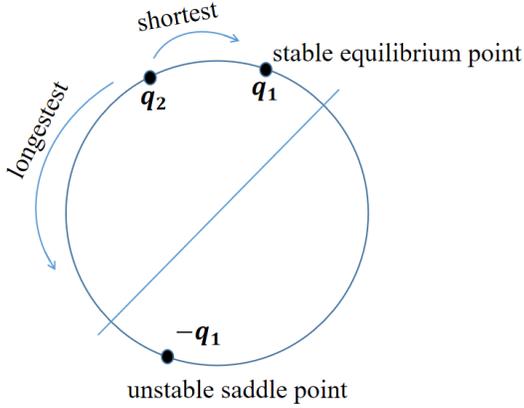}
\caption{Unit constraint sphere surface, where a point denotes a specific orientation and two antipodes ($q_{1}$ and $-q_{1}$) denote the same orientation. When $q_{2}$ goes to unstable saddle point $-q_{1}$ on $\mathbb{S}^{3}$, then it is close to the physical equilibrium on manifold $SO(3)$; however, it will leave the physical equilibrium on manifold $SO(3)$ and then come back the physical equilibrium, since it will leave the unstable saddle point $-q_{1}$ to the stable equilibrium point $q_{1}$ on $\mathbb{S}^{3}$.}\label{shortest}
\end{figure}

As shown above, as the minimal globally nonsingular parametrization, unit quaternions are often used to parametrize rotation matrices on $SO(3)$. Meanwhile, however, given a certain attitude, there exist actually two sets of unit quaternions describing the same orientation in terms of $\mathcal {R}(-q)=\mathcal {R}(q)$, due to the \emph{double covering} transformation from $SO(3)$ to state space $\mathbb{S}^{3}$ of unit quaternions \cite{2012tac[4]}. As a result, this will cause the rigid body to unnecessarily make a full rotation from the unstable saddle point to another stable equilibrium point, called unwinding phenomenon \cite{10.7, chat2009tac}. As shown in Fig. \ref{shortest}, the stable one could specify the orientation by the shortest single axis rotation, while the other unstable saddle needs to specify by the longest.

As analyzed in \cite[pg. 103-111]{2019Ehsan[27]}, the kinematic differential equation of unit quaternions is rigorously linear if the angular velocities in all three directions depend only on time, and more generally considered bilinear if the angular velocities in all three directions are themselves coordinates. In fact, only the once-redundant unit quaternions (1 parameter is redundant compared with the 3-parameter case) retain a singularity-free attitude representation, whose kinematic differential equations are also linear analogous to the rotation matrix. Those properties show that unit quaternions are attractive attitude representation in some topics including attitude estimation problems using filters or observers where the KDEs are linearized. However, the kinematic differential equations for all 3-parameter attitude representations are nonlinear, containing geometrical and/or mathematical (kinematical) singularities.

\emph{The analysis on unwinding phenomenon and Gimbal lock:}\\
\indent--Given unit quaternion $q=(\eta, \epsilon)=(\cos \frac{\Phi}{2},\vec{k}\sin \frac{\Phi}{2})\in \mathbb{S}^{3}$. That is, $\eta=\cos \frac{\Phi}{2}$. Following \cite[(3.73)\&(3.91a)]{2019Ehsan[27]}, we get $\cos\Phi=\frac{1}{2}(C_{11}+C_{22}+C_{33}-1)$ where $C_{ii}$ denotes three diagonal elements of DCM.\\
\indent--Based on \cite[pg. 86-95]{2019Ehsan[27]}, Gimbal lock occurs when the \emph{pitch} is rotated $90^{\circ}$ up or down ($\pm 90^{\circ}$) in pitch-yaw-roll angles, or when an \emph{inclination} angle of $0^{\circ}$ or $180^{\circ}$ in the symmetric Proper Euler angles. Gimbal lock would cause disastrous results if the rigid-body is in a steep ascent or dive, in a gimbal-based aerospace inertial navigation system. Next we analyze the two cases when Gimbal lock occurs.

For the asymmetric pitch-yaw-roll angles, we define $(\theta_{1}, \theta_{2}, \theta_{3}):=(\psi, \textcolor[rgb]{0.00,0.00,0.00}{\theta}, \phi)$. Then we get the DCM in terms of the asymmetric pitch-yaw-roll angles, i.e.,
\begin{equation}\label{ypr-1}
\begin{split}
[C]=\left[\begin{array}{ccc}
       C_{2}C_{1} & C_{2}S_{1}& -S_{2}\\
       S_{3}S_{2}C_{1}-C_{3}S_{1}& S_{3}S_{2}S_{1}+C_{3}C_{1} & S_{3}C_{2}\\
       C_{3}S_{2}C_{1}+S_{3}S_{1}& C_{3}S_{2}S_{1}-S_{3}C_{1}& C_{3}C_{2}
     \end{array}\right]
\end{split}
\end{equation}
where element $C_{2}$ denotes $\cos (\theta_{2})$, and others are similarly defined.
When Gimbal lock occurs, i.e., $\theta_{2}=\pm 90^{\circ}$, we have $C_{2}=0$ and $S_{2}=\pm 1$, then the DCM (\ref{ypr-1}) becomes
\begin{equation}\label{ypr-2}
\begin{split}
[C]&=\left[\begin{array}{ccc}
       0 & 0& \mp1\\
       \pm S_{3}C_{1}-C_{3}S_{1}& \pm S_{3}S_{1}+C_{3}C_{1} & 0\\
       \pm C_{3}C_{1}+S_{3}S_{1}& \pm C_{3}S_{1}-S_{3}C_{1}& 0
     \end{array}\right]  \\
     &= \left[\begin{array}{ccc}
       0 & 0& \mp1\\
       - S(\theta_{1}\mp \theta_{3})& C(\theta_{1}\mp \theta_{3}) & 0\\
       C(\theta_{1}\mp \theta_{3})& - S(\theta_{3}\mp \theta_{1})& 0
     \end{array}\right].
\end{split}
\end{equation}
That is, $C_{13}=\mp1$, $C_{11}=C_{33}=0$, $C_{22}=C(\theta_{1}\mp \theta_{3})$.

For the symmetric Proper Euler angles, we define $(\theta_{1}, \theta_{2}, \theta_{3}):=(\tilde{\psi}, \textcolor[rgb]{0.00,0.00,0.00}{\tilde{\theta}}, \tilde{\phi})$. Then we get the DCM in terms of the symmetric Proper Euler angles, i.e.,
\begin{equation}\label{noypr-1}
\begin{split}
[C]=\left[\begin{array}{ccc}
       C_{3}C_{1}-S_{3}C_{2}S_{1}& C_{3}S_{1}+S_{3}C_{2}C_{1}-& S_{3}S_{2}\\
        -S_{3}C_{1}-C_{3}C_{2}S_{1}&-S_{3}S_{1}+C_{3}C_{2}C_{1} & C_{3}S_{2}\\
        S_{2}S_{1} & -S_{2}C_{1}& C_{2}
     \end{array}\right]
\end{split}
\end{equation}
where element $C_{2}$ denotes $\cos (\theta_{2})$, and others are similarly defined.
For the Gimbal lock, i.e., $\theta_{2}=0^{\circ}~or~180^{\circ}$, we have $S_{2}=0$ and $C_{2}=\pm 1$, then the DCM (\ref{noypr-1}) becomes
\begin{equation}\label{noypr-2}
\begin{split}
[C]&=\left[\begin{array}{ccc}
       C_{3}C_{1}\mp S_{3}S_{1}& C_{3}S_{1}\pm S_{3}C_{1}-& 0\\
        -S_{3}C_{1}\mp C_{3}S_{1}&-S_{3}S_{1}\pm C_{3}C_{1} &0\\
        0 & 0& \pm1
     \end{array}\right]  \\
     &= \left[\begin{array}{ccc}
       C(\theta_{1}\pm \theta_{3})&  S(\theta_{3}\pm \theta_{1})& 0\\
        - S(\theta_{3}\pm\theta_{1})& -C(\theta_{1}\mp \theta_{3}) & 0\\
       0 & 0& \mp1\\
     \end{array}\right].
\end{split}
\end{equation}
That is, $C_{11}=C(\theta_{1}\pm \theta_{3})$, $C_{22}=-C(\theta_{1}\mp \theta_{3})$, $C_{33}=\mp1$.

\footnotetext [2]{https://blog.csdn.net/AndrewFan/article/details/60981437.}

\begin{figure}[!htb]
\centering
\includegraphics[width=2.8in,height=1.2in]{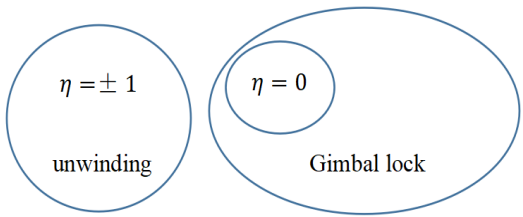}
\caption{Gimbal lock and unwinding phenomenon.}\label{unGL}
\end{figure}

Following \cite[(3.73)\&(3.91a)]{2019Ehsan[27]}, we get $\cos\Phi=\frac{1}{2}(C_{11}+C_{22}+C_{33}-1)$, so we do the following analysis using the asymmetric pitch-yaw-roll angles, since (\ref{noypr-2}) is more complicated than (\ref{ypr-2}). When Gimbal lock occurs in terms of (\ref{ypr-2}), we get $\cos\Phi=\frac{1}{2}(C_{11}+C_{22}+C_{33}-1)=\frac{1}{2}(0+C(\theta_{1}\mp \theta_{3})+0-1)=\frac{1}{2}(C(\theta_{1}\mp \theta_{3})-1)$. Then we do the following analysis:\\
\indent (a) if $C_{22}=1$, i.e., $\{1\}\subset C(\theta_{1}\mp \theta_{3})$ (like $\theta_{1}-\theta_{3}$=0), we obtain $\cos\Phi=\frac{1}{2}(1-1)=0$, so $\Phi=\frac{\pi}{2}$;\\
\indent (b) if $C_{22}=-1$, i.e., $\{-1\}\subset C(\theta_{1}\mp \theta_{3})$ (like $\theta_{1}+\theta_{3}=\pi$), we obtain $\cos\Phi=\frac{1}{2}(-1-1)=-1$, so $\Phi=\pi$, i.e., $\eta=\cos \frac{\Phi}{2}=0$.
In summary, Gimbal lock includes the case where $\eta=0$. However, Gimbal lock does not includes the cases where $\eta=\pm1$, as shown in the following analysis\footnotemark[3]:\\
\indent (c) when $\eta=\cos \frac{\Phi}{2}=\pm1$, then $\Phi=0~or~2\pi$, i.e., $\cos\Phi=1$, since $\cos\Phi=\frac{1}{2}(C_{11}+C_{22}+C_{33}-1)=\frac{1}{2}(0+C(\theta_{1}\mp \theta_{3})+0-1)=\frac{1}{2}(C(\theta_{1}\mp \theta_{3})-1)$, then it should hold $1=\frac{1}{2}(C(\theta_{1}\mp \theta_{3})-1)$, i.e., $C(\theta_{1}\mp \theta_{3})$ should be 3, which is impossible, resulting in a paradox.

As a result, Gimbal lock (i.e., when the plane's head is straight up or straight down as shown in Fig. \ref{GB}(b), corresponding to ``$\hbox{pitch}=\theta=\pm 90^{\circ}$'' or ``$\hbox{inclination}=\tilde{\theta}=0^{\circ}~or~180^{\circ}$'') includes the case where $\eta=0$, not including the cases where $\eta=\pm1$, as shown in Fig. \ref{unGL}. This is reasonable, since when the plane's head is straight up or straight down, we could only ensure the second rotation angle $\theta=\pm 90^{\circ}$ or $\tilde{\theta}=0^{\circ}~or~180^{\circ}$, but we could not ensure the first and third rotation angles, so that Gimbal lock may include many combinations of attitude angles.

\footnotetext [3]{Gimbal lock does not includes the cases where $\eta=\pm1$ when we do analysis using the symmetric Proper Euler angles, since Gimbal lock is the same singularity for all set of Euler angles, i.e., when the plane's head is straight up or straight down.}

As analyzed above, the essence of Gimbal lock is that we cannot control the attitude state where the pitch angle approaches to
$\pm 90^{\circ}$ (north/south pole) in the asymmetric Tait-Bryan angles. Fortunately, unite quaternions representation can overcome the difficulty by supporting spherical linear interpolation. However, because the covering is double ($q$ and $-q$ map to the same rotation), the rotation path may turn either the ``short way'' (less than $180^{\circ}$) or the ``long way'' (more than $180^{\circ}$), see Fig. \ref{shortest}. As a result, unwinding phenomenon arises, corresponding the cases where $\eta=\pm1$. To summarize, the unwinding phenomenon is caused when we ignore the two closed-loop equilibrium points $\mathcal {M}_{0}=\{q\in \mathbb{S}^{3}:\eta=\pm1\}$. That is, even when the Gimbal lock problem does not occur using unit quaternions, unwinding phenomenon follows. Therefore, in existing literature, they usually restrict the rotation angle to be in $[-\pi, \pi)$ and thus obtain the almost global attitude control for example in \cite{30abd}. Another method is applying the switching logic variable in order to derive the global attitude control, where both of the two equilibrium points $\eta=\pm1$ is globally asymptotically stable.

\subsection{(Modified) Rodrigues parameters}\label{section3.5}

Rodrigues parameters and modified Rodrigues parameters are also three-parameter representations derived from unit quaternions denoted as $q=(\eta, \epsilon^{T})\in \mathbb{S}^{3}$ where $\epsilon\in \mathbb{R}^{3}$. So Rodrigues parameters or modified Rodrigues parameters are also minimal. Similar to Eluer angles, the three-parameter sets of (modified) Rodrigues parameters can be regarded as certain embedded subsets of Euclidean space $\mathbb{R}^{3}$, so that one can apply the analysis methods suited for $\mathbb{R}^{3}$.

According to \cite[pg. 20]{abd2013shu}, we present the Rodrigues vector
\begin{equation}\label{MRP1}
\begin{split}
\rho:=\frac{1}{\eta} \epsilon=\vec{k} \tan(\textcolor[rgb]{0.00,0.00,0.00}{\Phi}/2),
\end{split}
\end{equation}
where $\vec{k}\in \mathbb{R}^{3}$ is an arbitrary unit-length vector and $\textcolor[rgb]{0.00,0.00,0.00}{\Phi}$ is the rotation angle. We call the three elements of $\rho$ the Rodrigues parameters.

As a minimal 3-parameter representation of attitude, rotations $\textcolor[rgb]{0.00,0.00,0.00}{\Phi}=\pm \pi$ do not make sense in (\ref{MRP1}). Hence, Rodrigues vectors could not represent rotation angles $\textcolor[rgb]{0.00,0.00,0.00}{\Phi}=\pm \pi$, corresponding to $\eta=0$ in terms of the analysis on connections between unwinding phenomenon and Gimbal lock in Section \ref{section3.2}. It can be checked from (\ref{MRP1}) that the small angle behavior of Rodrigues parameters could be more linear than that of any Euler angle representation, since $\tan(\textcolor[rgb]{0.00,0.00,0.00}{\Phi}/2)\approx \textcolor[rgb]{0.00,0.00,0.00}{\Phi}/2$ for small angle $\textcolor[rgb]{0.00,0.00,0.00}{\Phi}/2$. So linearizing Rodrigues parameters could nicely remove from the singularities $\textcolor[rgb]{0.00,0.00,0.00}{\Phi}=\pm \pi$.

Next, we present a modified representation called modified Rodrigues parameters \cite[pg. 20]{abd2013shu}, i.e.,
\begin{equation}\label{MRP2}
\begin{split}
\bar{\rho}:=\frac{1}{1+\eta} \epsilon=\vec{k} \tan(\textcolor[rgb]{0.00,0.00,0.00}{\Phi}/4),
\end{split}
\end{equation}
where $\vec{k}\in \mathbb{R}^{3}$ is an arbitrary unit-length vector and $\textcolor[rgb]{0.00,0.00,0.00}{\Phi}$ is the rotation angle. We call the three elements of $\rho$ the Rodrigues parameters.

As a minimal 3-parameter representation of attitude, rotations $\textcolor[rgb]{0.00,0.00,0.00}{\Phi}=\pm 2\pi$ corresponding to $\eta=-1$ do not make sense in (\ref{MRP2}). Hence, modified Rodrigues vectors could not represent rotation angles $\textcolor[rgb]{0.00,0.00,0.00}{\Phi}=\pm 2\pi$ and/or $\eta=-1$. That is, the
singularity in (\ref{MRP2}) has moved to $\pm 2\pi$ from $\pm \pi$ in the Rodrigues vector (\ref{MRP1}). So, except the complete rotation from the current orientation to the original orientation, modified Rodrigues parameters could describe any rotation angle.

\subsection{Comparisons}\label{section3.6}

As analyzed in the former subsections, each attitude representation has its strengths and weaknesses
compared to the others:

--As shown in Table \ref{unique}, 3-parameter representation includes Euler angles and (modified) Rodrigues parameters, which could neither globally nor uniquely define the attitude rotations. According to the analysis in Section \ref{section3.2} for Euler angles, the geometric singularity occurs at certain specific values of the second Euler rotation angle (pitch angle at $\pm 90^{\circ}$ and/or inclination at angle $0^{\circ}$ or $180^{\circ}$), so the Euler angles representation is suitable for vehicles which do not need the pitch angles (i.e., no vertical maneuvers), including land vehicles and ships \cite{Dobook}. Also, as analyzed in Section \ref{section3.5}, (modified) Rodrigues parameters, which are derived from unit quaternion, could provide a continuous and unique attitude rotation if we limit rotation angles between $-\pi$ and $\pi$ for Rodrigues parameters and between $0$ and $2\pi$ for modified Rodrigues parameters (i.e., rotation angle is less than $2\pi$), respectively. If there is no constraint on rotation angles, then geometrical singularity occurs at ($\pm2\pi$) $\pm\pi$ for (modified) Rodrigues parameters.

--As shown in Table \ref{unique}, 4-parameter representation includes quaternion and axis-angle, which could globally not uniquely define the attitude rotations. According to the analysis in Section \ref{section3.4} with unit quaternions significant advantages, unit quaternions are often used to represent the attitude on $SO(3)$ since the computational burden can be greatly reduced compared with rotation matrix.
However, the once-redundant unit-quaternion representation needs 4 parameters, which is an over-parameterization of the boundaryless compact manifold $SO(3)$, so that the transformation $q\mapsto SO(3)$ is a double covering map (i.e., two-to-one map). Therefore, there exist two equilibrium solutions in the unit quaternion space for each equilibrium solution on $SO(3)$, one is stable equilibrium solution while the other is unstable saddle point, causing unwinding phenomenon. More specifically, when trajectories starts near a desired
equilibrium solution on manifold $SO(3)$, then it may diverge on $SO(3)$ if the corresponding equilibrium solution on the quaternion space is the unstable saddle point, then it comes back to the same equilibrium solution on manifold $SO(3)$ after traveling a large distance.
According to the analysis on axis-angle in Section \ref{section3.3}, we know that axis-angle representation is similar to unit quaternion, i.e., globally but not uniquely, with similar constraints of double covering map.

\begin{figure*}
\centering
\includegraphics[width=6in,height=3.3in]{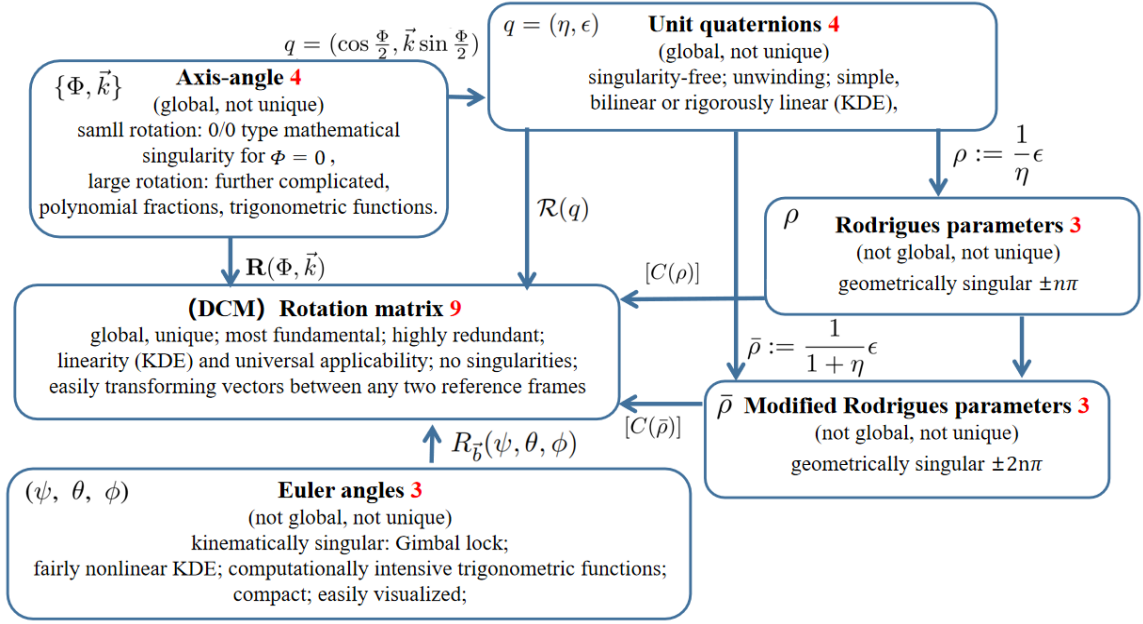}
\caption{Comparisons among the six kinds of attitude representations in Table \ref{unique}, where ``DCM'' denotes direction cosine matrix and ``KDE'' denotes kinematic differential equation.}\label{r6}
\end{figure*}

--In contrast, rotation matrix could provide a global and unique representation for all attitudes without any singularity. However, the 9 elements of the DCM are trigonometric functions, bringing more cumbersome computations, as analyzed in Section \ref{section3.1}. Compared with works for using rotation matrix (directly using the Lie group $SO(3)$) to avoid singularities, there is a weakness using unit quaternions, i.e., non-uniqueness. However, as presented in Table \ref{unique}, attitude rotations necessitate dealing with nine elements of the rotation matrix, and each element includes several
trigonometric functions making it harder to deal with. Therefore, as a global (non-singular) parameterization equivalent to rotation matrix, unit quaternion is simpler to deal with, compared with rotation matrix. Refer to Fig. \ref{r6} for the connections among those attitude representations.

\begin{figure}
\centering
\includegraphics[width=3in,height=2.4in]{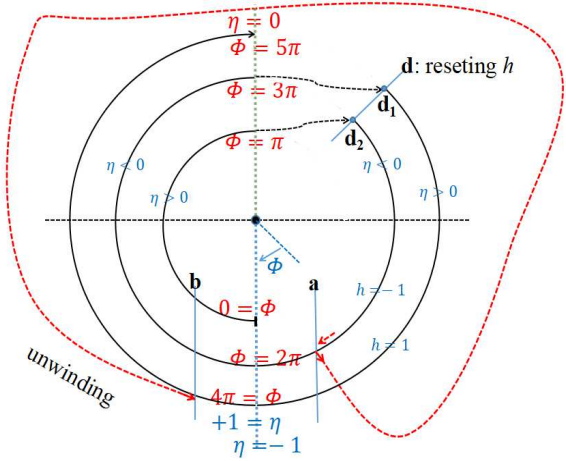}
\caption{Axis angle $\Phi$ and unit quaternion $q=(\eta, \epsilon)=(\cos \frac{\Phi}{2},\vec{k}\sin \frac{\Phi}{2})$: continuous control, discrete control and hybrid control. For $\eta=1$, we have $\Phi\in \{4n\pi\}=\{0, 4\pi, 8\pi, 12\pi, \ldots\}$ with $n=1,2,3,\ldots$; for $\eta=0$, we have $\Phi\in\{(1+4n)\pi, (3+4n)\pi\}=\{\pi, 3\pi, 5\pi, 7\pi, 9\pi, \ldots\}$; for $\eta=-1$, we have $\Phi\in \{2(2n+1)\pi\}=\{2\pi, 6\pi, 10\pi, \ldots\}$.}\label{3style}
\end{figure}

\section{Attitude Control }\label{section4.0}
This section analyzes unit quaternion based control laws and DCM based control laws for (almost) global control in Section \ref{section4.1}, \ref{section4.2} and \ref{section4.3} respectively, as well as robustness analysis in Section \ref{section4.4} and technique tools' analysis in Section \ref{section4.5}.

\subsection{Unit quaternion based control laws}\label{section4.1}

The non-uniqueness issue of unit quaternions: \\
\indent --Essential reason of non-uniqueness: Non-uniqueness is caused by the fact that the boundaryless
compact manifold $SO(3)$ is not diffeomorphic to any Euclidean space, so that the map $\mathcal {R}: \mathbb{S}^{3}\rightarrow SO(3)$ is double covering, i.e., $\mathcal {R}(-q)=\mathcal {R}(q)$ for any $q\in \mathbb{S}^{3}$. \\
\indent --Solution: Solving the non-uniqueness by considering the double covering issue and thus adopting the switching control law as in \cite{2011tac}, in order to avoid the unwinding phenomenon and realize the global attitude control.

Consider the case where $q=(\eta, \epsilon)=(\cos \frac{\Phi}{2},\vec{k}\sin \frac{\Phi}{2})\in \mathbb{S}^{3}$ for given axis-angle parameter $\{\Phi, \vec{k}\}$, with the two closed-loop equilibrium points $\mathcal {M}_{0}=\{q\in \mathbb{S}^{3}:\eta=\pm1\}$.

\emph{Unit-quaternion-based control for the kinematics: continuous control, discrete control and hybrid control.}

1) Continuous control: robustly but not globally controlled.\\
\indent --Consider the continuous controller $\omega=-\epsilon$, used to stabilize the equilibrium point $q=(1,0,0,0)$, then another equilibrium point $q=(-1,0,0,0)$ is unstable saddle point, i.e., the attitude is not globally stabilized. When $\Phi\rightarrow 2\pi$ clockwise as shown in Fig. \ref{3style}, it will suddenly rotate reversely from certain point $\textbf{a}$ nearby $2\pi$, tend to point $\textbf{b}$ nearby $\Phi=4\pi$ and finally be stabilized at stable equilibrium point $q=(1, 0,0,0)$ with $\Phi=4\pi$, i.e., unwinding phenomenon occurs. That is, for $\Phi\in\{2\pi, 6\pi, 10\pi, \ldots\}$, it will unnecessarily rotate $2\pi$ reversely and finally be stabilized at the stable equilibrium point $q=(1,0,0,0)$ , i.e., causing unwinding phenomenon as shown in Fig. \ref{3style}.

2) Discrete control: neither robustly nor globally controlled (almost globally).\\
\indent --Consider the discrete controller $\omega=-\eta\epsilon$, i.e., $\omega=-\epsilon$ is to stabilize the equilibrium point $q=(1,0,0,0)$ while $\omega=\epsilon$ is to stabilize the equilibrium point $q=(-1,0,0,0)$. When $\Phi\rightarrow 2\pi$ clockwise then it will cross certain point $\textbf{a}$ nearby $2\pi$ as shown in Fig. \ref{3style} and finally be stabilized using $\omega=-\epsilon$ at stable equilibrium point $q=(-1, 0,0,0)$ with $\Phi=2\pi$; When $\Phi\rightarrow 4\pi$ clockwise then it will cross certain point $\textbf{a}$ nearby $4\pi$ and finally be stabilized using $\omega=\epsilon$ at stable equilibrium point $q=(1, 0,0,0)$ with $\Phi=4\pi$. As a result, even if the state arrives around $\textbf{a}$, it will not rotate reversely, thus no unwinding phenomenon occurs. However, if the initial state starts from $\mathcal {M}=\{q\in \mathbb{S}^{3}:\eta=0\}$ (i.e., a nowhere dense set of zero Lebesgue measure) for $\Phi\in \{\pi, 3\pi, 5\pi, \ldots\}$ as shown in Fig. \ref{3style}, then it will take arbitrarily long time to the two closed-loop equilibrium points $\mathcal {M}_{0}=\{q\in \mathbb{S}^{3}:\eta=\pm1\}$, and even never achieve at $\mathcal {M}_{0}$. Consequently, the attitude is almost globally stabilized, not globally stabilized. For example, the controller $\omega=-\epsilon$ switches to $\omega=\epsilon$ at $\Phi=\pi$, i.e., the initial state for the equilibrium point $q=(-1, 0,0,0)$ is actually $\mathcal {M}=\{q\in \mathbb{S}^{3}:\eta=0\}$. Hence, every time when $h$ switches, then $\mathcal {M}=\{q\in \mathbb{S}^{3}:\eta=0\}$ is actually the initial state for the next stabilization round, i.e., every time $h$ switches, it will take arbitrarily long time to achieve the next stable state. Moreover, at $\Phi\in\{\pi, 3\pi, 5\pi, 7\pi, 9\pi, \ldots\}$ as shown in Fig. \ref{3style}, switching controller between $\omega=-\epsilon$ and $\omega=\epsilon$ is not robust, i.e., any noise may affect the measurement of $\eta$ causing chattering at the discontinuity.

3) Hybrid control: robustly and globally controlled.\\
\indent --Consider the discrete controller $\omega=-h\epsilon$ where $h$ is reset at $h\eta=-\delta$ for constant $0<\delta<1$, i.e., $\omega=-\epsilon$ is to stabilize the equilibrium point $q=(1,0,0,0)$ where $h=-1$ switches to $h=1$ at certain point $\textbf{d}_{1}$ for $\eta=(\geq)\delta$ as shown in Fig. \ref{3style}; while $\omega=\epsilon$ is to stabilize the equilibrium point $q=(-1,0,0,0)$ where $h=1$ switches to $h=-1$ at certain point $\textbf{d}_{2}$ for $\eta=(\leq)-\delta$ as shown in Fig. \ref{3style}. As a result, as analyzed in the discrete case, no unwinding phenomenon occurs. Similar to the discrete case where $h$ should switch at $\mathcal {M}=\{q\in \mathbb{S}^{3}:\eta=0\}$ (i.e., the North pole in Fig. \ref{3style}), but no robustness holds. So switching at $h\eta=-\delta$ (i.e., switching $h=-1$ to $h=1$ at $\eta\geq \delta$ and switching $h=1$ to $h=-1$ at $\eta\leq -\delta$) means that the switching point at original points $\mathcal {M}=\{q\in \mathbb{S}^{3}:\eta=0\}$ has been removed to $\textbf{d}$, thus robustness holds. Consequently, the attitude is globally stabilized.

\subsection{DCM based control laws}\label{section4.2}


To realize the robust global attitude control, \cite{2011tac} proposes the binary logic variable based quaternion hybrid control law (transforming the unstable saddle point to the stable equilibrium point), where the hysteresis half-width arranges a tradeoff between the hysteresis-induced inefficiency for avoiding unwinding phenomenon and the robustness to measurement noise. Differently, a class of synergistic potential functions based hybrid control could also overcome the topology constraints on $SO(3)$ (preventing all potential functions from arriving their unstable saddle points) and thus realize the robust global control using (globally and nonsingularly defined) DCM on $SO(3)$ \cite{TL2}, \cite{15abd}, \cite{8.1}, \cite{16abd}, \cite{8.2}.

In the synergistic potential functions based hybrid control, one needs to construct a family of synergistic potential functions usually based on the basic potential function--modified trace function--with the angular warping technique \cite{15abd}, \cite{16abd}, or a class of more intuitive and straightforward expelling configuration error functions \cite{TL2}.
Generally, all the hybrid attitude control laws consist in three term: a proportional term, a derivative term resulting form the angular velocity that provides damping, as well as a feedforward torque. Differently, the proportional term depends on the discretely changing gradient of potential functions in the potential functions based hybrid control \cite{8.1}, \cite{15abd}, while depends on the continuously changing state $\epsilon$ and binary logic variable $h$ \cite{2011tac}. For the proportional term depending on the discrete gradient of potential functions, one also needs to smooth the hybrid controller via applying a dynamic (continuous but nonsmooth) interpolation method between the switching potential-energy terms based on the backstepping method \cite{8.2} or based on the integral action directly on the proportional term \cite{15abd}, in order to produce a continuous torque command. The purpose is to response the potential caveat of the hybrid feedback, i.e., some unmodeled dynamics--flexible structures, sloshing fuel or overwhelm the available actuators--maybe excited by the discontinuities in the commanded torque \cite{8.1}, since the potential function is switched and those control-induced functions are to stabilize a unique desired point. There does not exist such problems in the binary logic variables $h\in\{-1,1\}$ based hybrid control, since there only exist two control paradigms, i.e., both of the two critical points $(1, 0,0,0)$ and $-(1,0,0,0)$ could be stabilized respectively by logic variables $h=1$ and $h=-1$.

In summary, comparisons between different hybrid control laws can be summarized as:
1) All of them deal with robust global attitude control, i.e., overcoming the topology constraints of boundaryless compact manifold $SO(3)$, based on configuration error functions (in \cite{TL2}) or synergistic potential functions (in \cite{15abd}), or binary logic variable based hybrid control (in \cite{2011tac}).
2) \cite{TL2} and \cite{15abd} apply the global and nonsingular rotation matrices without adopting any attitude parametrization. In \cite{TL2} and \cite{15abd}, the switching among synergistic functions implies the global attitude control, since all potential functions are prevented from arriving their unstable saddle points. So constructing a family of synergistic functions is challenging. \cite{2011tac} applies the unit quaternion parametrization, where the computational burden can be greatly reduced and the switching of a logic variable based control law implies the global attitude control (by transforming the unstable saddle point to a stable equilibrium point), but analyzing the robustness of the hybrid control law brings difficulties.

\subsection{Analysis on (almost) global control}\label{section4.3}



\begin{table*}[tbp]
\centering  
\begin{tabular}{|c|c|c|}
\hline
Attitude Representation & almost global control\footnotemark[4] & global control \\ \hline

Euler angles  &  ~~  & \cite{10.7[1]} via synthesizing a class of control laws \\ \hline

~ & ~ & \cite{renwei2008shu} restricting the rotation angle to be in $[0, \pi]$ \\
(Modified) Rodrigues parameters  & ~ &  \cite{10.2} via modifying control laws over a set of zero Lebesgue measure\\ \hline


~& ~ & \cite{2011tac, zddtcsI, zddtcns} with binary logic variables\\
Quaternions&  \cite{30abd}, \cite{10.6}   & \cite{31abd} with smooth control (rotation angle of the attitude error is different from $\frac{\pi}{2}$) \\ \hline
Axis-angle & ~~&~ \\  \hline
~ & ~~& \cite{TL9} by switching controllers \\
~ & \cite{TL5, TL10}& \cite{TL1, TL2} with multiple configuration error functions \\
Rotation matrix & \cite{TL1, TL2} with smooth control &  \cite{8.1, 8.2} and \cite{15abd} with a family of synergistic potential functions   \\
~ & ~& \cite{TL11} via globalizing the local stable manifold by the backward flow map \\ \hline 
\end{tabular}
 \caption{Properties of attitude representations II.}\label{almost}
\end{table*}

As shown in Table \ref{unique}, only DCM could globally and uniquely represent the attitude. However, this does not mean that the attitude system can achieve global control under rotation matrix (especially when the controller is continuous) such as \cite{TL1, TL2} with smooth control besides \cite{TL5, TL10}, since rotation matrix belongs to Lie group $SO(3)$ whose internal inherent topology (not contractible, boundaryless compact manifold, not diffeomorphic to any Euclidean space) completely excludes the existence of any continuous feedback control in order to achieve global attitude stabilization on $SO(3)$. As well, unit quaternion based continuous control law could not realize the global attitude control, either, such as \cite{30abd}. Hence, the internal inherent topology of $SO(3)$ means that any continuous control using any attitude representation (or directly DCM) is impossible to realize the global attitude stabilization on $SO(3)$.

As a result, no matter which parameter is used, i.e., unit quaternions, DCM or (modified) Rodrigues parameters, designing discontinuous control laws is usually required in order to realize the global control. Moreover, considering that purely discontinuous control laws are usually not robust to any measurements thus causing chattering as analyzed in the coming Section \ref{section4.4}, researchers usually design switching control laws (also called hybrid control laws), such as \cite{2011tac} with binary logic variables using unit quaternions, \cite{TL1, TL2} with multiple
configuration error functions using DCM, \cite{8.1, 8.2} and \cite{15abd} with a family of synergistic potential functions using DCM, \cite{TL9} by switching controllers using DCM where the attraction region for each control mode covers the configuration space almost globally.

However, it is not always the case. Besides the hybrid control laws, the following cases also bring the global control as shown in Table \ref{almost}: the case in \cite{TL11} via globalizing the local stable manifold by the backward
flow map using DCM; the case in \cite{mengziy2019auto} where only one equilibrium is considered using unit quaternion; the case in \cite{31abd} with smooth control using unit quaternions provided that the rotation angle of the attitude error is different from $\pi/2$; the case in \cite{10.5} by restricting control gains for global asymptotic stability proved by the Lyapunov method, where the designed nonlinear control law using unit quaternions is model-independent (moments of inertia is not involved) and gain matrices are symmetric positive definite. If the control gains are not restricted in \cite{10.5}, i.e., for any choice of control gains, then the desired equilibrium could only be locally asymptotically stabilized, instead of the global result. However, there is no systematic synthesis procedure choosing those control gains in such case \cite{10.5}, which could only be done by trial and error. In addition, for the modified Rodrigues parameters representation, modifying these control laws over a set of zero Lebesgue measure is feasible to realize the global asymptotic stabilization control over the whole manifold $SO(3)$, for example using an open-loop strategy applied over an arbitrarily small finite interval \cite{10.2}. As well, the modified Rodrigues parameters representation could realize the global attitude control by restricting the rotation angle to be in $[0, \pi]$ \cite{renwei2008shu}.

\footnotetext [4]{\cite{TL2} Consider an equilibrium of a dynamic system located
at the origin. The equilibrium is almost globally asymptotically stable, if it is asymptotically
stable and almost all trajectories converge to it, i.e., the set of
the initial states that do not asymptotically converge to the origin
has zero Lebesgue measure.}

\subsection{Analysis on robustness}\label{section4.4}

Generally, there exist two kinds of robustness, i.e., one is the nominal robustness with respect to small perturbations (including measurement noises/disturbances) for avoiding chattering, and the other is the general robustness against bounded \emph{unstructured} uncertainties in both the translational dynamics and the rotational dynamics.

For hybrid attitude control to achieve \emph{global} attitude control, a hysteresis-based switching algorithm is often designed where the nominal robustness holds for avoiding chattering like \cite{2011tac}, \cite{TL1}, \cite{TL2}, \cite{TL9}. Generally, for hybrid control systems, nominal robustness holds if the hybrid assumption conditions (i.e., Ass. 6.5 in \cite{2012teel}) are satisfied, like the analysis in \cite{2011tac}. The nominal robustness holds in \cite{10.5} to model errors and parametric uncertainties, since the designed nonlinear control law in \cite{10.5} is model-independent (moments of inertia is not involved) and gain matrices are symmetric positive definite. Despite one of the designed nonlinear control laws \cite{10.7} is also model-independent (i.e., nominal robustness holds), however, the control gains are scalars, not including matrices gains.

As well, the general robustness against \emph{unstructured} uncertainties implies that the tracking/stabilization errors are
uniformly ultimately bounded, for example, \cite{TL1, TL2} and \cite{TL5} apply an integral control term to achieve robust global
exponential/asymptotic stability in the presence of disturbances. In many works, the authors only consider one kind of robustness, for example \cite{2011tac}, \cite{8.1} and \cite{15abd} for nominal robustness, \cite{mengziy2019auto} for general robustness, while in \cite{13-14} and in most works done by \emph{Taeyoung Lee}, both kinds of robustness are simultaneously considered for the global attitude control, like \cite{TL1}, \cite{TL2}, \cite{TL3}, \cite{TL4}, \cite{TL9}.

Different from other works using Euler angles with kinematical singularities like \cite{10.7[1]} or quaternions with the ambiguities (causing the unwinding phenomenon) \cite{2011tac}, \cite{13-14}, the works of \emph{Taeyoung Lee} directly use the special Euclidean group $SO(3)$, which are intrinsic and coordinate-free, thereby completely avoiding any singularities and ambiguities, as well as those works using a family of ``synergistic'' potential functions \cite{8.1} and \cite{15abd} for realizing global attitude control. For the case where the special Euclidean group is used, classical trace functions \cite{TL10} or modified trace functions \cite{8.1} are usually designed as Lyapunov functions to derive the global attitude stability.



%

\subsection{Analysis on technique tools}\label{section4.5}

\footnotetext [5]{A nonlinear system is said to be autonomous if it does not
depend explicitly on time \cite{TL4barbalat}. }

Not all theoretical methods are available to all situations. For example, the LaSalle's/hybrid invariance principle, which is better applied to aysmptotic stability of certain set points for time-invariant (i.e., autonomous\footnotemark[5]) \cite[Pg. 1370]{10.7[20I]} or periodic equation \cite{10.1}, \cite{10.2}, \cite{2011tac}, \cite{8.1}, not including the time-varying (i.e, non-autonomous) systems \cite{10.4}. Other than LaSalle's/hybrid invariance principle, we may use cross-term-added Lyapunov function or Barbalat's lemma for the time-varying case \cite{TL4}, \cite{TL5}. Comparison principle can be applied into non-autonomous/autonomous cases \cite{2015tangauto},
\cite[Chapter IX]{10.7[20I]}. Matrosov' theorem can be applied in all circumstances \cite{5.2shs[38]}, i.e., time-(in)variant systems, nonperiodic/time-dependent systems, etc. In summary, one may use other methods for non-autonomous systems, for example Barbalat's lemma \cite[Pg. 105]{TL4barbalat}, Matrosov's theorem, comparison principle, etc.

Despite \cite{10.5} and \cite{10.7} both design model-independent quaternion control laws for attitude control systems. However, the former uses the LaSalle's theorem to derive the global asymptotic stability when there is no cross-term in the designed Lyapunov function; while the latter applies Barbalat's theorem to derive the global asymptotic stabilization control where a cross term is added in the designed Lyapunov function. Even if the designed Lyapunov function does not contain any cross terms, Barbalat's lemma is also feasible to derive the stability control for the unit-quaternion based autonomous attitude control systems \cite{30abd}, \cite{31abd}. In fact, there is no need to use LaSalle's invariance principle when the Lyapunov function contains a cross term like \cite{TL2, TL9}. The work in \cite{10.7[20I]} about robotic manipulators further explores why the LaSalle's invariance principle and Barbalat's lemma are not required, when cross-term is added in the designed Lyapunov function. In fact, in non-autonomous systems, there is no need to use the LaSalle's/hybrid invariance principle if one could find a Lyapunov function with strictly negative derivative. However, for the passive/dissipative systems (i.e., attitude systems), one have to apply the LaSalle's/hybrid invariance principle when the designed Lyapunov function is semi-negative, unless adding a cross term in the Lyapunov function \cite{10.7}.

\section{Conclusion and Discussion}\label{section6}
The attitude control problem is presented in this paper. A class of attitude representations are clarified and compared according to the parameter set and geometric/kinematicla singularities, including 9-parameter rotation matrix, 3-parameter Euler angles and (modified) Rodrigues parameters, 4-parameter axis-angle and unit quaternions. According to attitude representations, almost global and global attitude control can be achieved based on different control laws where the nominal and general robustness are analyzed, including continuous control laws (i.e., modifying control laws over a set of zero Lebesgue measure or restricting the rotation angle for the global control), discontinuous control laws without nominal robustness and hybrid control laws (i.e., switching controllers or multiple configuration error functions for the global control), etc. According to different attitude representations, technical tools are presented for deriving the (almost) global stabilization, i.e., the LaSalle's/hybrid invariance principle for aysmptotic stability for time-invariant or periodic equation, cross-term-added Lyapunov function or Barbalat's lemma for time-varying case, comparison principle for non-autonomous/autonomous case, Matrosov' theorem for all circumstance (the time-(in)variant systems, nonperiodic/time-dependent systems).

\end{document}